\documentclass[aps,prl,twocolumn,epsfig,showpacs]{revtex4}

\usepackage{graphicx}
\usepackage{epsfig}
\usepackage{dcolumn}
\usepackage{bm}
\usepackage{color}

\begin{document}

\title{Blistering of viscoelastic filaments}

\author{R.~Sattler$^{1,*}$, C.~Wagner$^{1}$, J.~Eggers$^{2}$} \affiliation{$^1$
Experimentalphysik, Universit\"at des Saarlandes, Postfach 151150,
66041 Saarbr\"ucken, Germany  \\
$^2$  School of Mathematics, University of Bristol, University
Walk, Bristol BS8 1TW United Kingdom    $^*$ r.sattler at
mx.uni-saarland.de}

\begin{abstract}
When a dilute polymer solution experiences capillary thinning, it
forms an almost uniformly cylindrical thread, which we study
experimentally. In the last stages of thinning, when polymers have
become fully stretched, the filament becomes prone to instabilities,
of which we describe two: A novel ``breathing'' instability,
originating from the edge of the filament, and a sinusoidal
instability in the interior, which ultimately gives rise to a
``blistering'' pattern of beads on the filament. We describe the
linear instability with a spatial resolution of 80 nm in the
disturbance amplitude. For
sufficiently high polymer concentrations, the filament eventually
separates out into a ``solid'' phase of entangled polymers,
connected by fluid beads. A solid polymer fiber of about $100$
nanometer thickness remains, which is essentially permanent.
\end{abstract}

 \maketitle \vskip2pc
When a drop falls from a faucet, surface tension drives the fluid
motion toward breakup in finite time, and a drop separates. This
pinch-off occurs in a localized fashion \cite{CCFF04}, and the
neighborhood of the point of breakup is described by a similarity
solution \cite{E93}. If however very small amounts of high molecular
weight polymer are added, an almost perfectly cylindrical thread is
formed instead \cite{Bazilevskii1981,
Amarouchene2001,Wagner2005,Clasen2006}. The reason is that wherever
there is a local decrease in radius, fluid elements are stretched,
and the polymers along with it. This will increase the extensional
viscosity of the fluid-polymer mixture \cite{Bird87}, and further
flow is inhibited, thus forming a uniform and stable filament.

For most of this paper, we produce a filament by placing a drop of
liquid between two solid plates, which are rapidly drawn apart
\cite{Anna2001}. (In a simple and educational version of this
experiment, a drop of saliva is placed between thumb and index
finger). A single filament forms between the plates, which thins as
surface tension drains fluid from the filament, and into
two roughly hemispherical reservoirs at the endplates.

\begin{figure}
\epsfig{file=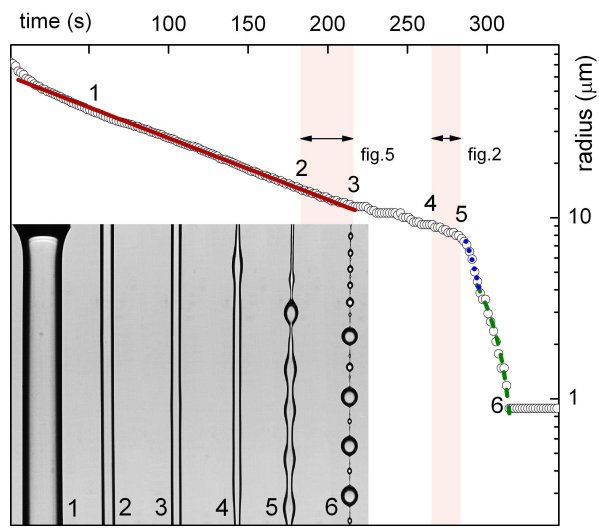, angle=0, width=0.9\linewidth}
\caption{Color online. The minimum radius $h_{min}$ as a function of
time. Not all data points are shown. Between number 1 and 3, the
curve is well described by $h_{min}(t)=h_{0}exp(-t/\tau)$ with $\tau
= 130\pm30ms$ (red, straight line). A plateau is reached between $3$
and $4$ which is associated with an instability near the endplates,
followed by rapid pinching. Between 5 and 6 the curve can be
approximated with two linear laws (dotted blue and dashed green lines). For
further explanation see text.} \label{fig1}
\end{figure}

The present paper addresses the later stages of the thinning of the
polymeric filament, when polymers have come close to their full
extension. Thus the mechanism that formerly used to stabilize the thread is
no longer effective, and tiny beads begin to appear on the filament,
(see Fig.~\ref{fig1}, images 5 and 6) \cite{movie}. We will refer to
this process central to the present study as ``blistering''. This
instability occurs when the filament is only $12 \mu m$ in radius,
requiring extreme spatial resolution. If the concentration of
polymer is above 1000 ppm, the filament can
become very long-lived compared to the timescale of a
dissolved polymer.

Theoretically, the period of exponential thinning
has recently been described within a long-wavelength description
\cite{Clasen2006}. Nonetheless, the full three-dimensional,
axisymmetric problem remains unsolved. The effect of finite polymer
extensibility has been studied numerically in \cite{Li2003}, once
more using a long-wavelength model. The filament is found to fail
near its end via a localized similarity solution, in contrast to
the much more complex scenario found here. The first clear experimental
description of blistering is found in \cite{Oliveira2005}, which
focuses on the later stages of the instability, in the course of
which droplets with a hierarchy of sizes are found. We first focus
on the {\it onset} of the blistering instability, for which widely diverging
theoretical explanations have been expressed in the past
\cite{Chang1999,Li2003,Oliveira2005,Clasen2006}.

Our experiments were performed with aqueous solutions of
polyethylene oxide (PEO) of molecular weights $M_{W}=1-8 \times
10^{6} amu$ and concentrations $0.010 \leq c \leq 0.2$ weight
percent. While the formation of a filament and its subsequent
instability could be observed well below the overlap concentration
\cite{Bird87} of polymers, we focused on higher concentrations, as
the process is slower and easier to observe. Our reference
system has $M_{W}=4 \times 10^{6} amu$ and $c = 0.2$ weight percent,
with $c_{ov}=0.07$ weight percent the overlap concentration. The
samples were characterized with a Thermo Haake MARS rheometer using
cone plate geometries. The zero shear viscosity was
$\eta_{0}\approx50mPas$ and in the range of shear rates $0.1
\leq\dot{\gamma}\leq 2000$ shear thinning was present down to a
value of $\eta_{\infty}\approx4mPas$. The surface tension was
determined by the pendant drop method to $\gamma\sim60.9mN/m$
\cite{Oliveira2006,rheo}.

We used a capillary breakup device similar to the one described in
\cite{Oliveira2005,Rodd2005}. This setup is also commercially
available to measure the extensional rheology of suspensions (CaBER,
Thermo Fisher Scientific, Karlsruhe, Germany). To ensure maximum
reproducibility, we used the following protocol: plates of diameter
$d=2 mm$ were held at a distance of $l=2.5mm$ for the purpose of
relaxation for several seconds. Then the plates are drawn to
$l=3.5mm$ within 40ms, only slightly exceeding the limit at which a
capillary bridge of the solvent experiences a Rayleigh-Plateau
instability and breakup. The thinning process is observed with an
IDT X-Vision X3 digital high-speed video camera with Nikon
Microscope objectives of up to $20\times$ magnification. At the
highest magnification, the diffraction limited resolution is
$0.6\mu m$, and depth of field is $5\mu m$. A Halogen back-light allows frame
rates up to 6000 frames per second and exposition times down to
$10\mu s$.

Figure~\ref{fig1} shows a typical recording of the thread radius
$h_{min}(t)$ in the cylindric region in a semi-logarithmic plot. For plug flow in a
cylindrical filament, the elongation rate is determined from
$\dot{\epsilon} = -2 d\ln h/dt$, thus $\dot{\epsilon}$ is constant
for most of the filament thinning, which follows an exponential law
(the regime between 1 and 2 in Fig.~\ref{fig1})
\cite{Amarouchene2001}. The axial stress $\sigma_{zz}$ supported by
the polymers balances the increasing capillary pressure $\gamma /
h_{min}$, which means that the extensional viscosity $\eta_E \equiv
\sigma_{zz}/\dot{\epsilon}= \gamma/(h_{min}\dot{\epsilon})$ also
increases exponentially. At 3 (see
Fig.~\ref{fig1}) the thread radius first has a plateau, and then
thinning accelerates again. The reason for the acceleration is that
the polymers have almost reached their maximum extension, so their
extensional viscosity can no longer increase. But this means that
$\dot{\epsilon}$ has to increase rapidly, implying a steep increase
in the slope of $\log h(t)$, as seen in Fig.~\ref{fig1}. Once
$\eta_E$ has reached a plateau, which we estimate at $h_{min}\approx
12 \mu m$ to be $\eta_{E}\left({12\mu m}\right)\approx 330 Pa s$,
the filament behaves essentially like a Newtonian fluid
\cite{Li2003}, and is thus subject to a capillary instability
\cite{E97}, as confirmed below. Note that this value of the
extensional viscosity corresponds to an increase by 5 orders of
magnitude over $\eta_{water} = 10^{-3} Pas$ of the solvent.

\begin{figure}
\epsfig{file=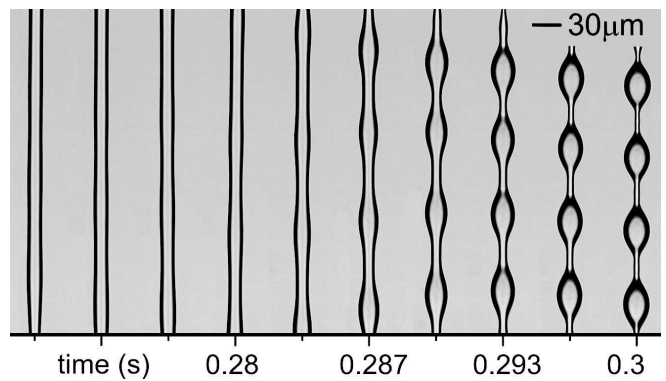, angle=0, width=1\linewidth}
\caption{Growth of a sinusoidal instability of the viscoelastic
filament that develops into a group of droplets on the thinning
filament. The spacing of the pictures is $300^{-1}s$. The time window
between 4 and 5 in Fig.~\ref{fig1} is represented by the
images up to $0.287s$ showing the range of exponential growth.
} \label{fig2}
\end{figure}

\begin{figure} \epsfig{file=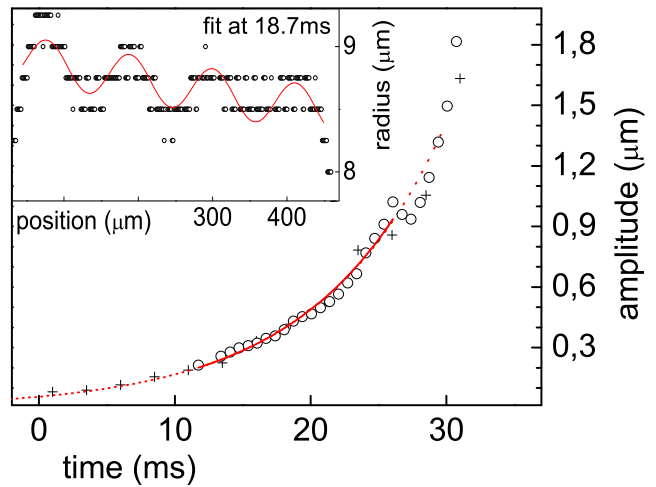, angle=0,width=1\linewidth}
\caption{Main: The growth of the amplitude of a sinusoidal surface
deformation on a filament of radius $R_0=10\mu m$. The origin of the
time axis has been shifted relatively to Fig.~\ref{fig1}. Circles
and crosses are experimental data of two different runs, from
the latter we were able to detect amplitudes as low as 80 nm.
The straight and the dotted line are exponential fits,
giving an inverse growth rate of $\omega = 9.3\pm 0.1
ms$. Inset: The sinusoidal surface deformation at $t=18.7ms$. Points
are experimental data and the solid line is a fit with a sine and a
linear offset. The selected wavelength is $\lambda/R_{0} = 12 \pm
0.9$. } \label{fig3}
\end{figure}

Before we describe the novel instability that occurs between 2 and 3 in
Fig.~\ref{fig1}), which is localized near the endplates, we concentrate
on the subsequent spatially uniform, {\it linear}
instability which is shown in Fig.~\ref{fig2} (between 4 and 5 in
Fig.~\ref{fig1}). At first, no oscillations are visible
on the images of Fig.~\ref{fig2}; however, as seen in
Fig.~\ref{fig3}, we are able to resolve perturbations down to an
amplitude of $A = 80 n m$, corresponding to significant
super-resolution \cite{H95}. This is done by fitting the profile with
a sine function with wavenumber, phase and amplitude as free
parameters over many wavelengths (see inset). The algorithm converged
down to the stated maximum resolution.
The last four pictures of Fig.~\ref{fig2}
show the beginning of the non-linear stages of the instability,
finally leading to the formation of smaller secondary droplets
\cite{Oliveira2005,Oliveira2006}. In the main panel of
Fig.\ref{fig3} we plot the growth of the sinusoidal approximation over time, an
example of which is shown in the inset. Over more than a decade, the
growth is very well described by an exponential, providing a clear
signature of a {\it linear instability}, which develops uniformly in
space.

>From a fit to the exponential, we find an inverse growth rate of
$1/\omega = 9.3\pm0.1ms$. Linear stability of a viscous fluid thread
\cite{E97} predicts $\omega=\gamma/(6R_{0}\eta_{eff})$, which gives
an estimated extensional viscosity of $\eta_{eff} =9Pas \pm 2$, more
than one order of magnitude smaller than the extensional viscosity
$\eta_{E}\left({12\mu m}\right)$ estimated above. At the same time,
we are able to fit - as expected \cite{E97} - a {\it linear} law
$h_{min} = -0.44\times 10^{-3} m/s \Delta t$  in the range $8\mu m >
h_{min} > 4\mu m$. Comparing to the law $h_{min} = 0.07
\gamma/\eta_{eff} \Delta t$ for viscous pinching \cite{Rothert2003},
this gives $\eta_{eff} = 10 Pa s$, which agrees nicely. We do not
have a ready explanation for the discrepancy between $\eta_{eff}$
and $\eta_{E}\left({12\mu m}\right)$. Among possible explanations
are non-uniformities in the polymer concentration (see below), and
transient relaxation of polymer stresses during the plateau between
3 and 4 in Fig.~\ref{fig1}, when there is no flow. A kink in the
linear shrinking behavior at $h_{min}\sim 3.8 \mu m$ toward a less
steep slope of $-0.17\times 10^{-3} m/s$ could either be seen
as a first indication for the onset of a draining process discussed
below, or as as a transition from the viscous to the
inertial-viscous pinch-off regime\cite{Rothert2003}.

\begin{figure}
\epsfig{file=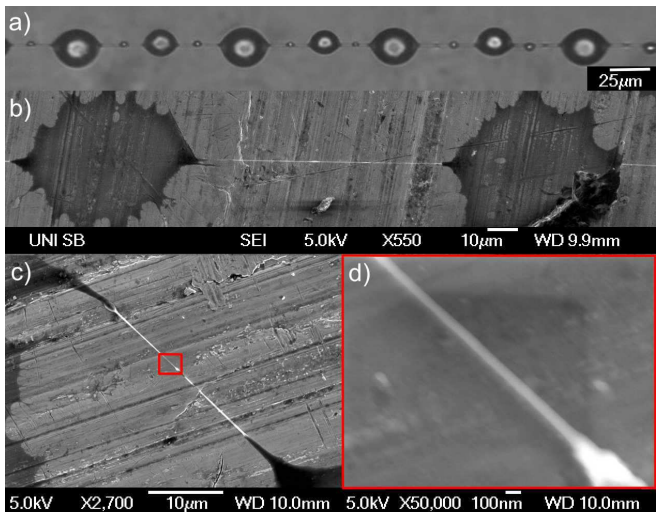, angle=0, width=1\linewidth}
\caption{a) The final state of the filament. Beads are formed
off-center relative to the thread. b) Scanning Electron Microscopy
image of two beads, connected by a thread (intermediate
resolution). The structure was caught and dried upon the substrate
c) Another example of the structure, the red box
indicating a closeup at high magnification shown in d). The diameter
of the fiber can be as small as $70nm$.} \label{fig4}
\end{figure}

The formation of successive generations of beads has already been
studied extensively \cite{Oliveira2005,Oliveira2006}; we focus on
the very final stages of the thinning process, when the formation of
new beads has come to rest (6 in Fig.~\ref{fig1}). If the polymer
concentration was greater than 1000 ppm, the filament connecting two
beads {\it never} breaks, and a pattern as shown in Fig.\ref{fig4}a)
is formed. What is remarkable is that most beads are off-center with
respect to the filament. Comparison with the problem of fluid drops
on a fiber \cite{Carroll86} shows that there must be a {\it finite}
contact angle between the drops and the filament for such a symmetry
breaking to occur. In other words, the thin filament must have
formed a (solid) phase different from that of the drops.

To confirm this idea, we produced the Scanning-Electron-Microscopy
(SEM) images shown in Fig.~\ref{fig4}. Object slides were pulled
quickly through the liquid bridge, before the accumulated elastic
stress would lead it to retract into a single droplet. Panel b)
shows two remnants of two droplets being connected by a persistent
thin thread. Increased magnification (panels c) and d)) allowed us
to estimate the diameter of the fiber as $75-150nm$. Assuming a
constant polymer concentration of the solution, and taking for the
fiber the density of PEO, the amount of polymer in such a fiber
equals a fluid diameter of $3\mu m$, thereby representing a lower
bound for the onset of the concentration process that leads to this
solid fiber, but which is likely to start earlier. Our physical
picture is that polymers become entangled, while solvent drains from
the filament, leading to even higher polymer concentration and
increased entanglement. Further evidence for this concentration
process was already found in \cite{Sattler2007}, where birefringence
measurements were performed to examine molecular conformations in
the break-up process. We can {\it exclude} evaporation to be a
factor in the formation of solid fibers, based on our estimates of
evaporation rates, as well as preliminary experiments in a two-fluid
system.

\begin{figure}
\epsfig{file=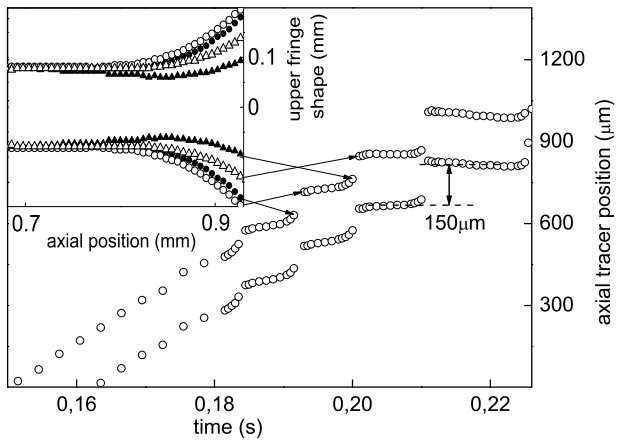, angle=0, width=1\linewidth}
\caption{Trajectories of two bubbles in the filament; they move in
parallel, indicating uniform flow inside the filament. First the
bubbles are convected by the extensional flow produced by the
thinning. Then their position undergoes a sequence of plateaus,
associated with consecutive contractions and relaxations - the
``breathing'' at the very end of the filament, shown in the inset.
The breathing correspondents to the red shaded region between 2 and
3 in Fig.~\ref{fig1}. At 3, the bubbles run out of the field of
view.} \label{fig5}
\end{figure}

Finally we would like to describe the transition regime (between 2
and 3 in Fig.~\ref{fig1}) where an instability of the homogenous
elongational flow originates from the boundary, and is reminiscent
of phenomena reported in \cite{Chang1999}. The transition region at
the edge of the filament was described in detail in
\cite{Clasen2006} for the case that polymers are far from stretched.
At full stretch, the transition region constricts (as seen in the
last shape in the inset of Fig.~\ref{fig5} (full triangles)), thus
inhibiting the flow out of the filament. The most sensitive probe
for the flow are tiny bubbles inside the filament, whose
trajectories are shown in Fig.~\ref{fig5}. During the constriction
phase, the flow stops, and the tracer positions form a plateau. In
the absence of flow, polymers relax inside the filament, and elastic
stresses are eventually no longer able to sustain the capillary
pressure. Fluid from the filament empties into the end cap, causing a
sudden flow, which appears as a jump in bubble position in
Fig.~\ref{fig5}. The process repeats itself periodically, on a
timescale that increases from step to step, but which is of
the same order of magnitude as the polymer relaxation time.

The height of the final jump gives a characteristic length scale of
$150 \mu m$, which is comparable to the wavelengths of periodic
disturbances on the filament (Fig.\ref{fig2}). In principle, each of
the plateaus shown in Fig.~\ref{fig5} should result in a
corresponding plateau in $h_{min}$. However, the plateaus are too
small to be resolved, apart from the last two between 3 and 4 in
Fig.~\ref{fig1}. At the end of the plateaus the filament shrinks
again by draining liquid into the reservoir at the endplates
or into a large bead that typically forms in the middle of the
filament, see \cite{movie}.

In conclusion, we have demonstrated three key phenomena:
\\
\noindent (i) Between 2 and 3 in Fig.~\ref{fig1}, the trumpet-shaped
transition region connecting the filament to the reservoirs
constricts periodically, interrupting the flow. \\
\noindent (ii) The blistering instability of the
filament, which leads to beads, is a linear capillary instability
(between 4 and 5 in Fig.~\ref{fig1}). As the polymers reach full
stretch, their contribution is once more Newtonian, but with a
viscosity that is many times that of the unstretched state. \\
\noindent (iii) Using electron microscopy, we provide evidence that
the filament remains intact in its latest stages, because polymer
strands become sufficiently concentrated to become solid-like. The
most compelling evidence for this fact is that the contact angle
between the thread and the beads sitting on it becomes {\it finite}.
Existing theoretical models are clearly inadequate in addressing
this polymer behavior at full stretch.

Acknowledgments: We thank J\"{o}rg Schmauch for the SEM images. This
work was supported by the DFG-Project WA 1336, the Royal Society and
Thermo Haake.

\end{document}